\documentclass[aps,prl,twocolumn,groupedaddress]{revtex4}
\usepackage{amsmath, amsfonts, amsthm, amssymb}
\usepackage[T1]{fontenc}
\usepackage[latin1]{inputenc}
\usepackage{calrsfs}
\usepackage{graphicx}
\usepackage{appendix}
\usepackage{natbib}
\usepackage{enumerate}
\usepackage{threeparttable}
\usepackage{setspace}
\usepackage{tabularx}
\usepackage{float}

\begin{document}

\title{Self-interaction correction in the LDA+U method}


\author{Dong-Kyun Seo}
\affiliation{Department of Chemistry and Biochemistry, Arizona State University, Tempe, AZ 85287-1604}


\date{\today}
\begin{abstract}
  We present one inherent shortcoming of the LDA+U method in respect of its self-interaction correction of the LDA. By reexamining the mean-field approximation on the Hubbard energy in the Hartree-Fock form, we have derived a new expression for the ``double-counting'' energy which led to a more reasonable self-interaction correction in an alternative LDA+U scheme. For the bulk Gd metal, the new scheme resulted in electronic properties more consistent with experiments in comparison to the LDA and the existing LDA+U schemes.  
\end{abstract}


\maketitle

\newcommand{\bfr}{\textbf{r}} 
\newcommand{\bfrp}{\textbf{r}'}

\newcommand{\rhor}{\rho(\bfr)} 
\newcommand{\rhop}{\rho(\bfrp)}

\newcommand{\rhoup}{\rho_{\uparrow}(\bfr)} 
\newcommand{\rhodn}{\rho_{\downarrow}(\bfr)}

\newcommand{\rhos}{\rho_{S}(\bfr)} 
\newcommand{\rhosp}{\rho_{S}(\bfrp)}

\newcommand{\rhoo}{\rho^{0}(\bfr)} 
\newcommand{\rhoi}{\rho_{m\sigma}^{0}(\bfr)}
\newcommand{\rhoij}{\rho_{ij\sigma}^{0}(\bfr)}

\newcommand{\psiis}{\psi_{m\sigma}(\bfr)} 
\newcommand{\psijs}{\psi_{j\sigma}(\bfr)}
\newcommand{\psijsp}{\psi_{j\sigma'}(\bfr)}
\newcommand{\psiiup}{\psi_{i\uparrow}(\bfr)} 
\newcommand{\psijup}{\psi_{j\uparrow}(\bfr)} 
\newcommand{\psiidn}{\psi_{i\downarrow}(\bfr)}
\newcommand{\psijdn}{\psi_{j\downarrow}(\bfr)}

\newcommand{\psijspo}{\psi_{j\sigma'}^{0}(\bfr)}
\newcommand{\psiiupo}{\psi_{i\uparrow}^{0}(\bfr)} 
\newcommand{\psijupo}{\psi_{j\uparrow}^{0}(\bfr)} 
\newcommand{\psiidno}{\psi_{i\downarrow}^{0}(\bfr)}
\newcommand{\psijdno}{\psi_{j\downarrow}^{0}(\bfr)}
\newcommand{\psiio}{\psi_{i}^{0}(\bfr)}
\newcommand{\psijo}{\psi_{j}^{0}(\bfr)}
\newcommand{\psipo}{\psi_{p}^{0}(\bfr)}
\newcommand{\psiqo}{\psi_{q}^{0}(\bfr)}
\newcommand{\psiiostar}{\psi_{i}^{0*}(\bfr)}
\newcommand{\psijostar}{\psi_{j}^{0*}(\bfr)}

\newcommand{\eis}{\epsilon_{m\sigma}}
\newcommand{\eiup}{\epsilon_{i\uparrow}}
\newcommand{\eidn}{\epsilon_{i\downarrow}}

\newcommand{\eiupo}{\epsilon_{i\uparrow}^{0}}
\newcommand{\eidno}{\epsilon_{i\downarrow}^{0}}
\newcommand{\eio}{\epsilon_{m}^{0}}

\newcommand{\nis}{n_{m\sigma}}
\newcommand{\niup}{n_{i\uparrow}}
\newcommand{\nidn}{n_{i\downarrow}}

\newcommand{\niupo}{n_{i\uparrow}^{0}}
\newcommand{\nidno}{n_{i\downarrow}^{0}}
\newcommand{\niso}{n_{m\sigma}^{0}}

\newcommand{\kinop}{-\frac{{\nabla^{2}}}{2}}
\newcommand{\vext}{v_{ne}}

\newcommand{\Erep}{\frac{1}{2}\int\!\!\!\!\int{\frac{{\rho(\bfr)\rho(\bfrp)}}{|\bfr-\bfrp|}d\bfrp d\bfr}}

\newcommand{\dr}{d\bfr} 
\newcommand{\drp}{d\bfrp}

Since its advent more than two decades ago~\cite{AZA:1991,ASKCS:1993}, the LDA+U method has remained as one of the most efficient methods that describe the strong correlations of electrons confined in an atomic region. (Throughout this paper LDA includes its generalization to spin-polarized systems.) When applied to the transition metal and rare-earth metal compounds, the LDA+U method gives a significant improvement compared with the LDA not only for excited-state properties but also for ground-state properties. However, further applications of the method may require its improvement. For example, it has been pointed out that weakly-correlated systems are not properly described by the LDA+U and that a correct prescription of the ``double-counting'' energy may be essential~\cite{PMCL:2003}. For strongly-correlated systems, there can exist drastic discrepancies in the locations of excited Hubbard states in comparison to the self-interaction-corrected LDA (SIC-LDA) calculations~\cite{SSTSW:1999}, which warrants a closer look at the self-interaction correction incorporated in the LDA+U. Herein, we present a shortcoming associated with its self-intereaction correction to the LDA, and provide an alternative LDA+U scheme with a new double-counting energy expression that does not bear the same problem, producing the results consistent with the SIC-LDA calculations.
 
The LDA+U total energy functional takes the form 
\begin{eqnarray}
E^{total}[\rho_{\sigma}, \hat{n}] &=&  E_{LDA}[\rho_{\sigma}] + E^{U}(\hat{n}) 
\label{E^total}
\; ,
\end{eqnarray}
where $E_{LDA}[\rho_{\sigma}]$ is the usual local spin-density functional of the total electron spin densities ($\rho_{\sigma}(\bfr)$, $\sigma = \uparrow, \downarrow$). The correction term to the LDA energy is given as
\begin{eqnarray}
E^{U}(\hat{n})  = E^{ee}(\hat{n}) - E_{LDA}^{dc}(\hat{n})
\label{E^U}
\; .
\end{eqnarray}
$E^{ee}$ is an electron-electron interaction energy of correlated $d$($f$) electrons in orbitals $\{\phi_{m}\}$, typically described by the multiband Hubbard model in the (unrestricted) Hartree-Fock (HF) approximation with a density matrix $\hat{n}$~\cite{SLP:1999}. $E_{LDA}^{dc}$ is the double-counting term that accounts approximately for the LDA-type energy of the correlated electrons. When electrons are in $\{\phi_{m}\}$, they feel the extra potential, $v_{m \sigma} ^{U}$, from $E^{U}(\hat{n})$ in addition to the LDA potential. The effective potential for those correlated electrons is given as
\begin{eqnarray}
v_{m \sigma} ^{eff}(\bfr)
                   = v_{ne}(\bfr) + v_{H}(\bfr)  + v_{xc \sigma}(\bfr) + v_{m \sigma} ^{U}
\label{v_msigma^eff}
\; .
\end{eqnarray}

\begin{figure}[h]
\centering
\includegraphics{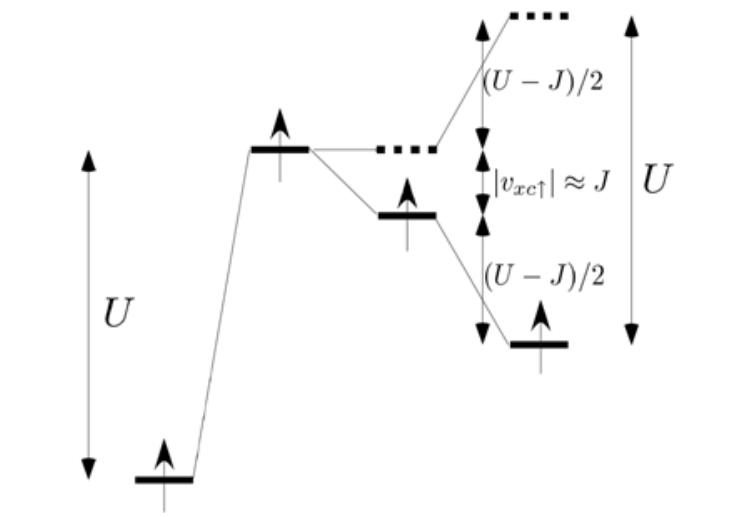}
\caption{\label{fig1} $1s$ spin-orbital energy changes of a hydrogen atom by subsequent inclusions of $v_{H} = U$, $\{ v_{xc \uparrow} \approx -J$,  $v_{xc \downarrow} \approx 0 \}$ and $v_{1s \sigma}^{\tiny{\textrm{AL}}} = \mp \frac{1}{2}(U-J)$ in the AL scheme. The dashed levels are the unoccupied $\downarrow$-spin energy level after the exchange splitting.}
\end{figure}

Depending on the way the $E_{LDA}^{dc}$ expression is derived, the current LDA+U methods are classified into two schemes, AMF (around mean field)~\cite{AZA:1991} and AL (atomic limit)~\cite{ASKCS:1993}. A drawback of the AMF scheme has been recognized, as it provides no effect on half-filled fully spin-polarized systems. The AL scheme does not have such a problem and hence is a preferred method for half-filled electron systems. However, we may recognize its shortcoming when we consider its $v_{m \sigma} ^{U}$ potential,  
\begin{eqnarray}
v_{m \sigma}^{\tiny{\textrm{AL}}}(\bfr) = -(U - J) \left( {n_{m \sigma} } - \frac{1}{2} \right)
\label{v^AL}
\; .
\end{eqnarray}
An illustrating case is an isolated hydrogen atom, a half-filled system.  With $N = 1$, $n_{1s \uparrow} = 1$ and $n_{1s \downarrow} = 0$, $v_{m \sigma} ^{eff}(\bfr)$ of Eq.~\ref{v_msigma^eff} is given as
\begin{eqnarray}
v_{1s \sigma}^{eff}(\bfr) &=& v_{ne}(\bfr) + U  + v_{xc \sigma}(\bfr)
                                 \mp \frac{U - J}{2} 
\label{v^AL-Hatom}
\; ,
\end{eqnarray}
where the $+$ and $-$ signs are for the $1s$ $\uparrow$- and $\downarrow$-spin orbitals of the hydrogen atom.
 Figure 1 shows how the $1s$ spin-orbitals change their energy, as $v_{H} = U$, $\{ v_{xc \uparrow} \approx -J $,  $v_{xc \downarrow} \approx 0 \}$ and $v_{1s \sigma}^{\tiny{\textrm{AL}}} = \mp \frac{1}{2}(U-J)$ are subsequently added.  While the $v_{1s \sigma}^{\tiny{\textrm{AL}}}$ clearly corrects the LDA term, $v_{xc \uparrow} \approx -J$, it does not bring the $\uparrow$-spin orbital back to the original position. In practice, the $(U - J)$ term in $v_{1s \sigma}^{\tiny{\textrm{AL}}}$ may be parametrized into $U^{\prime}$.  However, the $U^{\prime}$ value that completes the correction of $v_{H}$ in Fig. 1 will provide an erroneously large exchange splitting energy (on-site repulsion energy, in this case). This problem can be removed in an alternative LDA+U scheme proposed in the following.

We start with the electron-electron interaction energy within the HF approximation by using a diagonalized density matrix $\hat{n}$~\cite{AZA:1991,ASKCS:1993}:
\begin{eqnarray}
E_{HF} &=& \frac{1}{2} \sum_{m, m^{\prime}, \sigma} {U_{m m^{\prime}} n_{m \sigma} n_{m^{\prime} -\sigma}} 
                        \nonumber \\
            & &        + \frac{1}{2} \sum_{m \neq m^{\prime}, m^{\prime}, \sigma} 
                            {(U_{m m^{\prime}} - J_{m m^{\prime}}) n_{m \sigma} n_{m^{\prime} \sigma}}
\label{E_HF}
\; .
\end{eqnarray}
Eq.~\ref{E_HF}  is reexpressed to collect all Coulomb interaction terms in the Hartree energy component $E_{H}$ at the beginning:
\begin{eqnarray}
E_{HF} &=&    \frac{1}{2} \sum_{m, m^{\prime}, \sigma} { U_{m m^{\prime}}  
                                      (n_{m \sigma} n_{m^{\prime} -\sigma} 
                                      + n_{m \sigma} n_{m^{\prime} \sigma})}
                                      \nonumber \\
            & &
                        - \frac{1}{2} \sum_{m, \sigma} {U_{mm}n_{m \sigma}^2}
                        - \frac{1}{2} \sum_{m \neq m^{\prime}, m^{\prime}, \sigma} {J_{m m^{\prime}} n_{m \sigma} n_{m^{\prime} \sigma}} 
                       \nonumber \\
                  &\approx& E_{H}
                        - \frac{\check{U}}{2} \sum_{m, \sigma} {n_{m \sigma}^2}
                        - \frac{\check{J}}{2} \sum_{m \neq m^{\prime}, m^{\prime}, \sigma} {n_{m \sigma} n_{m^{\prime} \sigma}} 
\label{E_HF-alt}
\;  .
\end{eqnarray}
The $\check{U}$ and $\check{J}$ parameters are by definition averaged ``self-orbital'' Coulomb (or exchange) integral and ``pair-orbital'' exchange integral, respectively, for which a rotational invariance is assumed.  The $E_{H}$ term is often approximated as $\frac{1}{2} UN^{2}$ under the rotational invariance, where $U$ is the usual average Coulomb integral. For a more rigorous treatment, however, we leave $E_{H}$ as it is without the approximation. Alternatively, averaging of all the Coulomb (exchange) integrals gives the $U$ ($J$) parameter in the conventional approximate expression for the $E_{HF}$ in Eq.~\ref{E_HF}, however, not with $\frac{1}{2} UN^{2}$ but with $E_{H}$:
\begin{eqnarray}
E_{HF} &\approx&  \frac{1}{2} \sum_{m, m^{\prime}, \sigma} { U_{m m^{\prime}}  
                                      (n_{m \sigma} n_{m^{\prime} -\sigma} 
                                      + n_{m \sigma} n_{m^{\prime} \sigma})}
                                      \nonumber \\
            & &
                        - \frac{U}{2} \sum_{m, m^{\prime}, \sigma} {n_{m \sigma} n_{m^{\prime} \sigma}}
                        + \frac{U-J}{2} \sum_{m \neq m^{\prime}, m^{\prime}, \sigma} {n_{m \sigma} n_{m^{\prime} \sigma}} 
                       \nonumber \\
                  &=& E_{H}
                        - \frac{U - J}{2} \sum_{m, \sigma} {n_{m \sigma}^2}
                        - \frac{J}{2} \sum_{\sigma} {N_{\sigma}^{2}} 
\label{E_HF-org}
\;  .
\end{eqnarray}

Following the mean-field (MF) treatment in the AMF scheme~\cite{AZA:1991,PMCL:2003}, we obtain the LDA-type energy corresponding to Eq.~\ref{E_HF-alt} upon replacing all $n_{m \sigma}$'s by equal occupation numbers, $\frac{N_{\sigma}}{2l+1}$, among the $(2l+1)$-degenerate orbitals:
\begin{eqnarray}
E_{HF}^{\tiny{\textrm{MF}}} &\approx& E_{H}
                            - \frac{\check{U}^{\tiny{\textrm{MF}}} + 2l\check{J}^{\tiny{\textrm{MF}}}}{2(2l+1)}
                                \sum_{\sigma} {N_{\sigma}^2}
\label{E_HF^MF}
\;  .
\end{eqnarray}
The MF treatment may also affect numerical values of the $\check{U}$ and $\check{J}$ parameters, and hence the parameters are designated in Eq.~\ref{E_HF^MF} as $\check{U}^{\tiny{\textrm{MF}}}$ and $\check{J}^{\tiny{\textrm{MF}}}$, respectively. From Eq.~\ref{E_HF^MF}, the spin-polarization energy can be estimated for an open-shell system with an $M$ number of unpaired electrons by noting $ \sum_{\sigma} {N_{\sigma}^2} = \frac{1}{2}(N^{2} + M^{2})$:
\begin{eqnarray}
E_{sp}^{\tiny{\textrm{MF}}} &\approx& - \frac{\check{U}^{\tiny{\textrm{MF}}}+2l\check{J}^{\tiny{\textrm{MF}}}}{4(2l+1)} M^2
\label{E_sp^MF}
\; .
\end{eqnarray}
The quadratic dependence of the spin-polarization energy on the spin magnetic moment $M$ is consistent with the previous studies within LDA~\cite{SBJ:1993}.  In particular, our recent work has shown \textit{quantitatively}~\cite{Seo:2006} that within a second-order perturbational theory, the corresponding LDA expression is given for isolated atoms as
\begin{eqnarray}
E_{sp}^{LDA} &\approx& - \frac{U^{xc}+2lJ^{xc}}{4(2l+1)} M^2
\label{E_sp^LDA}
\; ,
\end{eqnarray}
in which the $U^{xc}$ and $J^{xc}$ are defined as self-orbital and pair-orbital exchange-correlation interaction energies, respectively. They define the $atomic$ exchange parameter~\cite{Seo:2006},
\begin{eqnarray}
I_{ex} = \frac{U^{xc}+2lJ^{xc}}{2l+1} 
\label{Iex}
\; ,
\end{eqnarray}
which is analogous to the Stoner exchange parameter of bulk metals. The two different sets of exchange parameters, calculated in LDA, are numerically identical for the $3d$ elements~\cite{Seo:2006}, which is expected to hold for the $4f$ elements as well. 

From the equivalency of the expressions in Eqs.~\ref{E_sp^MF} and \ref{E_sp^LDA}, we find the following relationship in replacement of Eq.~\ref{E_HF^MF}:
\begin{eqnarray}
E_{HF}^{\tiny{\textrm{MF}}} &\approx& E_{H} 
                        - \frac{I_{ex}}{2} \sum_{\sigma} {N_{\sigma}^2}
\label{E_HF^MF-app}
\; .
\end{eqnarray}
The second term in Eq.~\ref{E_HF^MF-app} can be considered approximately as an LDA exchange-correlation energy for the correlated electrons.  Because $I_{ex}$ is usually identified as the average exchange integral $J$~\cite{AZA:1991}, therefore, we conveniently replace the former by the latter in Eq.~\ref{E_HF^MF-app}. The final expression serves as $E_{LDA}^{dc}$ in our new LDA+U scheme:
\begin{eqnarray}
E_{LDA}^{dc} &=& E_{H} 
                        - \frac{J}{2} \sum_{\sigma} {N_{\sigma}^2}
\label{E^dc}
\; .
\end{eqnarray}
Inserting Eqs.~\ref{E_HF-org} and \ref{E^dc} in Eq.~\ref{E^U} results in 
\begin{eqnarray}
E^{U} &=&  - \frac{U-J}{2} \sum_{m, \sigma} {n_{m \sigma}^2}
\label{E^U-app}
\; .
\end{eqnarray}
It is recognized that the Hartree terms in $E_{HF}$ and $E_{LDA}^{dc}$ precisely cancel out each other in Eq.~\ref{E^U-app}. 

Accordingly, the corresponding potential for the electron of the index $n_{m \sigma}$ is obtained as the first-order derivative of $E^{U}$ with respect to $n_{m \sigma}$:
\begin{eqnarray}
v_{m \sigma}^{U} &=& - (U - J) n_{m \sigma}
\label{v_U}
\; .
\end{eqnarray}
When compared with $v_{m\sigma}^{\tiny{\textrm{AL}}}$ in Eq.~\ref{v^AL}, the new potential does not contain the extra term, $\frac{1}{2}(U-J)$, which uniformly raises the one-electron potential for all the $d$($f$) orbitals. 
Figure 2 illustrates the spin-orbital energy changes when $v_{m \sigma}^{U}$ in Eq.~\ref{v_U} is used for a hydrogen atom.  In comparison to the one-electron potential in Figure 1, the self-interaction correction in $v_{m \sigma}^{U}$ is complete in Figure 2 and the on-site repulsion energy is expressed correctly at the same time.  

\begin{figure}[h]
\centering
\includegraphics{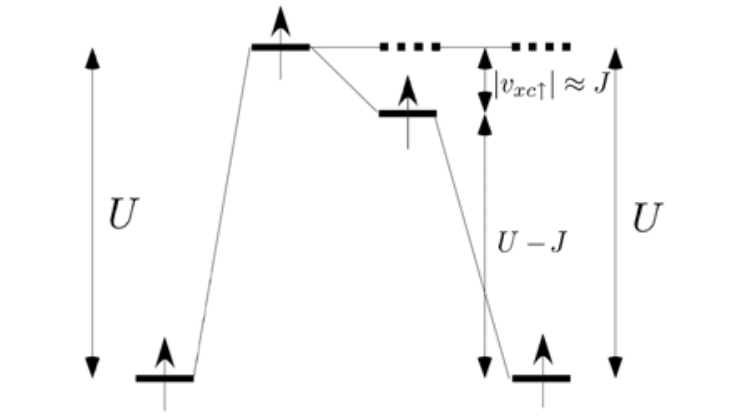}
\caption{\label{fig2} $1s$ spin-orbital energy changes of a hydrogen atom by subsequent inclusions of $v_{H} = U$, $\{ v_{xc \uparrow} \approx -J$,  $v_{xc \downarrow} \approx 0 \}$ and $\{ v_{1s \uparrow}^{U} = -(U - J)$, $v_{1s \downarrow}^{U} = 0 \}$ in the new LDA+U. The dashed levels are the unoccupied $\downarrow$-spin energy level after the exchange splitting.}
\end{figure}

In reality, the total electronic energy is calculated from the following expression~\cite{SLP:1999,EKC:2003}:
\begin{eqnarray}
E^{total}[\rho_{\sigma}, \hat{n}] &=& E_{LDA}(\epsilon_{i\sigma}^{LDA+U}) - E^{U}(\hat{n}) 
\label{E^total-alt}
\; .
\end{eqnarray}
$E_{LDA}(\epsilon_{i\sigma}^{LDA+U})$ is the total energy component that is obtained from the LDA expression by using the Kohn-Sham orbital energies, $\epsilon_{i\sigma}^{LDA+U}$, calculated from self-consistent LDA+U procedures.

 The new LDA+U scheme was implemented in FPLO (Full Potential Local Orbital) program (version 5)~\cite{KE:1999,EKC:2003} and was tested for the bulk Gd metal. A detailed analysis of the LDA results on the bulk Gd has been presented upon employment of the linearized augmented-plane-wave (LAPW) method in a scalar relativistic approximation~\cite{Singh:1991}. The unoccupied spin-minority $4f$ states are centered just 0.5 eV above the Fermi energy, and the consequent hybridization of the $4f$ states with the delocalized $5d$ states explains the complicated Fermi surfaces observed in experiments and also the appreciable contribution of the $4f$ electrons at the Fermi energy. The low-lying spin-minority $4f$ states have been observed later in the electronic structures of lanthanide metals and their compounds calculated with the SIC-LDA method, which accounts for the observed valence states across the lanthanide metal series~\cite{SSTSW:1999}. However, the LDA fails to predict the FM ground state of the bulk Gd. It was shown that LDA+U (the AL scheme) yields a correct FM ground state and this result has been attributed to an upward shift of the spin-minority $4f$ states from the Fermi energy in the LDA+U energy structure~\cite{HALSA:1995,SLP:1999}.

\begin{figure}[h]
\centering
\includegraphics{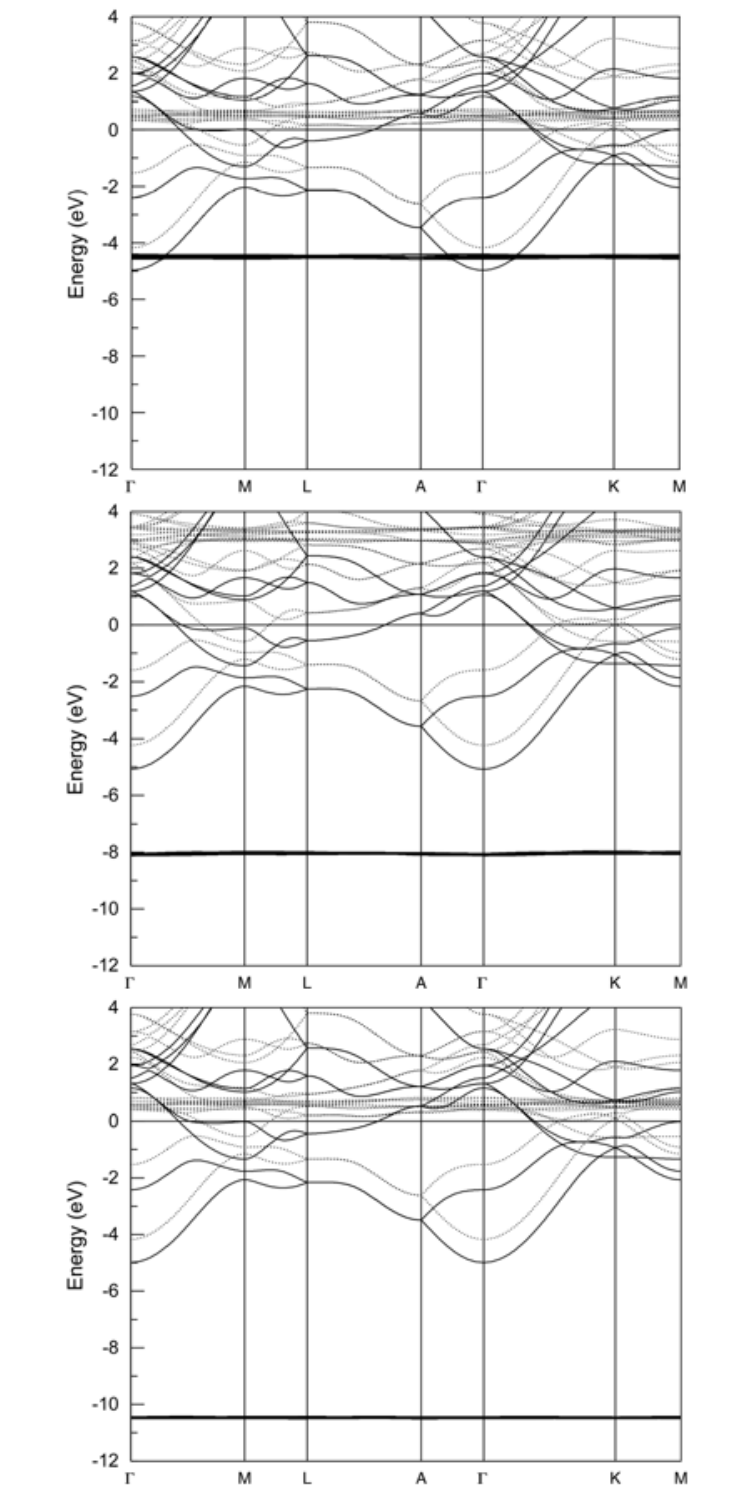}
\caption{\label{fig3} the electronic band structures of the FM Gd metal from the LDA (top), the LDA+U(AL)(middle) and new LDA+U (bottom) methods at the experimental structure.}
\end{figure}
 
 In our LDA+U calculations of the bulk Gd, we employed the Perdew-Wang LDA exchange-correlation functional~\cite{PW:1992}. For its $hcp$ structure, we chose the experimental lattice constants with $a =$ 6.8662 a.u. and $c/a =$ 1.587 used in the previous LDA+U(AL) LAPW calculations~\cite{SLP:1999}. $U =$ 6.7 and $J =$ 0.7 eV were used for the $4f$ orbitals ($F_{0} =$ 6.70, $F_{2} =$ 8.34, $F_{4} =$ 5.57,  and $F_{6} =$ 4.13 eV)~\cite{HALSA:1995,SLP:1999}. For the FM (AFM) state, 133 (372) $k$-points in the irreducible part of the hexagonal BZ were used, with Gaussian smearing. The compression parameters were optimized separately for the FM and AFM states for $4f$, ($5s$, $5p$), $5d$, $6s$ and $6p$ numerical atomic orbitals.  

Figure 3 shows the FM electronic structures of the bulk Gd calculated by LDA, LDA+U(AL) and new LDA+U in this work. The LDA result is in an excellent agreement with the previous LAPW calculation~\cite{Singh:1991}. The $4f$ states are separated into two manifolds by an exchange splitting of 5.0 eV and thus $I_{ex} = 0.71$ ($=\frac{5.0}{7}$) eV~\cite{Seo:2006}, conforming to the $J$ value employed for the LDA+U calculations~\cite{SLP:1999}. In comparison to the LDA, the LDA+U(AL) band structure shows a large shift of spin-minority $4f$ states by 2.5 eV, while the spin-majority $4f$ states are shifted downwards by 3.5 eV. The effective exchange splitting is larger than the LDA result by 6.0 eV, which is consistent with the previous LDA+U(AL) LAPW calculations~\cite{SLP:1999}.  In contrast to the LDA+U(AL), however, the new LDA+U scheme does not change the location of the spin-minority $4f$ states, and this leaves the band structure essentially unaffected near the Fermi energy. The close proximity of those $4f$ state around the Fermi energy is consistent with the observations from the SIC-LDA energy band structures of lanthanide metals and their compounds~\cite{SSTSW:1999}. The effective exchange splitting is again 6.0 eV larger than the LDA, which is solely due to the downward shift of the spin-majority $4f$ states.  

The spin magnetic moment and density of states for FM and AFM Gd are listed in Table I for the three calculations. Both of the spin magnetic moments from LDA and the new LDA+U results show a good agreement with experiment. The densities of states from the LDA and the new LDA+U are much closer to the experiment in comparison to the LDA+U(AL) result. Importantly, the new LDA+U predicts the correct FM ground state, although the stabilization against the AFM state is not as large as what we obtained from the LDA+U(AL). The stability of the FM ground state in the new LDA+U results indicates that the upward shift of the spin-minority $4f$ states may not be the origin of the stable FM state, in contrast to the previous suggestion~\cite{HALSA:1995,SLP:1999}.  

In summary, we have proposed a new LDA+U scheme that unerringly corrects the self interaction in the LDA. It is based on an MF expression of the sum of the exchange integrals and its connection to the LDA spin-polarization energy of atoms. The new method provided improved ground-state properties of the bulk Gd, while placing the upper Hubbard $4f$ states in the energy region found as in the SIC-LDA results.  Systematic studies on various systems will reveal in the future how the new method fares. Its close relationship to the existing LDA+U schemes allows easy implementation in other band structure calculation codes as well.  By employing an LDA exchange-correlation energy of the correlated electrons in the $E_{LDA}^{dc}$, the new scheme may eventually allow \textit{parameter-free} LDA+U calculations because the $E_{H}$ term in the $E_{LDA}^{dc}$ in our scheme is exact and becomes cancelled out in the $E^{U}$. 

This work was supported by the National Science Foundation through the author's CAREER Award (DMR - Contract No. 0239837) and also by his Camille Dreyfus Teacher-Scholar Award from the Camille and Henry Dreyfus Foundation. 

\small
\begin{spacing}{1}
\begin{threeparttable}
\caption{\label{tab1}
The spin magnetic moments (in $\mu_{B}$) and density of states at the Fermi energy (in (Ry atom)$^{-1}$) for FM and AFM states of the $hcp$ Gd from different calculation methods (AL: LDA+U(AL); SIC: the new LDA+U in this work). 
}
\centering
\begin{tabularx}{86mm}{ m{16mm}  X X X X}  \hline\hline  
\;              &LDA           & AL             & SIC      & Expt.\\ \hline 
FM             & 7.610       & 7.739        & 7.657        & 7.63$^{a}$      \\
\;               & 24.17       & 14.75        & 19.23        & 21.35$^{b}$       \\
AFM           & 7.384       & 7.628        & 7.439        &         \\ 
\;               & 22.09       & 17.85        & 21.20        &         \\ 
$\Delta E^{c}$  & $-$18.9            & 37.1                    & 9.5                       &          \\
\hline\hline
\end{tabularx}

\begin{tablenotes}
\item[\textit{a}] Reference~\cite{RCMMMJJ:1975}. 
\item[\textit{b}] Reference~\cite{WLJJ:1974}.
\item[\textit{c}] $E$(AFM) $-$ $E$(FM) in meV/atom.
\end{tablenotes}
\end{threeparttable}

\end{spacing}
\normalsize



\end{document}